\documentclass[a4paper,12pt]{article}

\usepackage{amsmath,amssymb,amsthm}
\usepackage[totalwidth=16.5cm,totalheight=23.5cm]{geometry}
\usepackage{graphicx}
\usepackage{comment}

\newtheorem{theorem}{Theorem}

\newcommand{\R}{{\mathbb R}}
\newcommand{\C}{{\mathbb C}}

\newcommand{\su}{\mathfrak {su}}

\newcommand{\be}{\begin{eqnarray}}
\newcommand{\ee}{\end{eqnarray}}

\def\beq{\begin{equation}}
\def\eeq{\end{equation}}
\def\ber{\begin{eqnarray}}
\def\eer{\end{eqnarray}}

\def\det{{\rm det\,}}

\begin{document}
 \pagestyle{plain}
\title{A 4D gravity theory and $G_2$-holonomy manifolds}
\author{Yannick Herfray${}^{(1),(2)}$, Kirill Krasnov${}^{(1)}$, Carlos Scarinci${}^{(1)}$ and Yuri Shtanov${}^{(3)}$\\ {}\\
{\small \it ${}^{(1)}$School of Mathematical Sciences, University of Nottingham, NG7 2RD,
UK}
\\
{\small \it ${}^{(2)}$ Laboratoire de Physique, ENS de Lyon,  46 all\'ee d'Italie, F-69364 Lyon Cedex 07, France}
\\
{\small \it ${}^{(3)}$Bogolyubov Institute for Theoretical Physics, 14-b Metrologichna
St., Kiev 03680, Ukraine} } 
\maketitle
\begin{abstract}\noindent
Bryant and Salamon gave a construction of metrics of $G_2$ holonomy on the total space of
the bundle of anti-self-dual (ASD) 2-forms over a 4-dimensional self-dual Einstein
manifold. We generalise it by considering the total space of an ${\rm SO}(3)$ bundle
(with fibers $\R^3$) over a 4-dimensional base, with a connection on this bundle. We make
essentially the same ansatz for the calibrating 3-form, but use the curvature 2-forms
instead of the ASD ones. We show that the resulting 3-form defines a metric of $G_2$
holonomy if the connection satisfies a certain second-order PDE. This is 
exactly the same PDE that arises as the field equation of a certain 4-dimensional gravity
theory formulated as a diffeomorphism-invariant theory of ${\rm SO}(3)$ connections.
Thus, every solution of this 4-dimensional gravity theory can be lifted to a
$G_2$-holonomy metric. Unlike all previously known constructions, the theory that we lift
to 7 dimensions is not topological. Thus, our construction should give rise to many new
metrics of $G_2$ holonomy. We describe several examples that are of cohomogeneity one on
the base.
\end{abstract}

\section{Introduction}

The history of $G_2$-geometry is almost as old as that of the exceptional Lie group $G_2$
itself, see \cite{History} for a nice exposition. The existence of metrics of $G_2$
holonomy was proven in \cite{Bryant}. This paper also gave a construction of the first
explicit example. Several more examples, among them complete, were constructed in
\cite{BS}. The first compact examples where obtained in \cite{Joyce}. 
More local examples can be obtained by evolving 6-dimensional ${\rm SU}(3)$
structures, see \cite{Hitchin}. These examples, as well as many other things, are
reviewed in \cite{Tour}. Metrics of $G_2$ holonomy are of importance in physics as
providing the internal geometries for compactification of M-theory down to 4 space-time
dimensions, while preserving supersymmetry. A nice mathematical exposition of this aspect
of $G_2$ geometry is given in \cite{Witt}.

Our interest in $G_2$ geometry is motivated by the fact that, as we explain in this
paper, solutions of certain 4D gravity theory can be lifted to $G_2$-holonomy metrics.
The gravity theory in question is {\it not\/} General Relativity, but rather a certain
other theory whose existence can be seen by reformulating 4D gravity as a diffeomorphism
invariant theory of ${\rm SO}(3)$ connections, as was described in \cite{Krasnov:2011up,
Krasnov:2011pp}, and explained from a more mathematical perspective in
\cite{Fine:2013qta}. Once 4-dimensional General Relativity is reformulated in the
language of connections, one finds that there is not one, but an infinite parameter
family of theories all resembling GR in their properties. The $G_2$-holonomy lift that we
describe in this paper singles out one of them, and it is distinct from GR. We describe
this theory in details in the main text.

A suggestion as to the existence of a link between some theory in 7 dimensions (referred
to as topological M-theory) and theories of gravity in lower dimensions was made in
\cite{Dijkgraaf:2004te}. That paper reinterpreted the constructions \cite{BS} of 7D
metrics of $G_2$ holonomy from constant curvature metrics in 3D and self-dual Einstein
metrics in 4D as giving evidence (among other things) for the existence of such a link.
The construction of the present paper is similar in spirit, but we present a much
stronger evidence linking 4D and 7D structures. Thus, our construction lifts any solution
of a certain 4D gravity theory with local degrees of freedom to a $G_2$ metric. The main
difference with the previous examples is that the theory that one is able to lift to 7D
is no longer topological. We find this result to be interesting as it interprets the
full-fledged 4D gravity as a dimensional reduction of a theory of differential forms in 7
dimensions.

We now formulate the main result of this paper. Let $A^i$, $i=1,2,3$, be an ${\rm SO}(3)$ connection 
on a 4-dimensional manifold $M$, and let $F^i = dA^i + (1/2)\epsilon^{ijk} A^j\wedge A^k$ be its curvature 2-form. 
Then, fixing an arbitrary volume form $v$ on $M$, define a $3\times 3$ symmetric matrix $X^{ij}$ by the relation
$F^i\wedge F^j= -2 X^{ij} v$. We will call a connection {\it definite} if $X^{ij}$ is a definite matrix, i.e. all eigenvalues have the same sign. The factor  of $2$ in the definition of the matrix $X^{ij}$ is introduced for the future convenience. 
The minus sign in the  same definition has to do with our later usage of anti-self-dual two-forms rather than self-dual ones. 

Let $E$ be an associated vector bundle over $M$ with 3-dimensional fibers. We consider the following 3-form on the total space of $E$:
\be\label{omega}
\Omega = \frac{1}{6}(1+\sigma y^2)^{-3/4} \epsilon^{ijk} d_A y^i \wedge d_A y^j \wedge
d_A y^k + 2\sigma(1+\sigma y^2)^{1/4} d_A y^i \wedge F^i.
\ee
Here $y^i$ are coordinates in the fiber, and $d_A$ is the covariant derivative with respect to $A$, and $\sigma=\pm 1$ is the sign of the connection to be defined below.

\begin{theorem} \label{theo}
If $A$ is a definite connection satisfying the second order PDE:
\be\label{det-eqs}
d_A \left[ (\det X)^{1/3} X^{-1} F \right] =0\, ,
\ee
then the 3-form (\ref{omega}) is stable, closed ($d\Omega=0$) and co-closed ($d
{}^*\Omega=0$), and hence defines a metric of $G_2$ holonomy.  This metric is of
Riemannian signature, and is complete (in the fiber direction) for $\sigma=+1$.
\end{theorem}

{\bf Remark.} The object $X^{-1}$ in (\ref{det-eqs}) is the symmetric matrix inverse to $X$, and 
$\det X$ is the determinant of $X$. Note that the expression under the covariant derivative in (\ref{det-eqs}) is of
homogeneity degree zero in $X$; therefore, equation \eqref{det-eqs} does not depend on
the particular choice of the orientation form $v$ used to define $X$. The sign $\sigma$ of a 
definite connection is defined in Sec.~\ref{sec:natural} below.
\bigskip

As we shall explain below, equation (\ref{det-eqs}) arises as the Euler--Lagrange
equation of a certain diffeomorphism-invariant theory of connections on $M$. Thus, the
theorem states that every solution of this theory can be lifted to a $G_2$-holonomy
metric in the total space of the bundle. As we shall also explain below, when the matrix
$X^{ij}\sim \delta^{ij}$, the connection $A$ is the anti-self-dual part of the
Levi-Civita connection on a self-dual Einstein manifold. In this case, equation
(\ref{det-eqs}) is satisfied automatically as it reduces to the Bianchi identity for the
curvature. The construction in the above theorem in this case reduces to that described
in \cite{BS}.

We also note that the metric in the total space $E$ of the bundle defined by the form
(\ref{omega}) induces a metric on the base $M$. This metric turns out to be in the
conformal class of
\be \label{urbanc}
g_F (\xi,\eta)\sim \epsilon^{ijk} i_\xi F^i \wedge i_\eta F^j \wedge F^k / v,
\ee
where $\xi$ and $\eta$ are vector fields tangent to the base. The volume form of the
metric induced on the base is a constant multiple of
\be\label{volume}
v_F = \left( \det X \right)^{1/3} v \, ,
\ee
where $v$ is the orientation form used to define the matrix $X$. It is easy to see
that the expression on the right-hand-side of \eqref{volume} does not depend on the
particular choice of the orientation form $v$, hence, is well-defined. The action
functional for the theory that gives rise to (\ref{det-eqs}) is just the total volume of
the space as computed using the volume form (\ref{volume}).

Another way to describe the relation between a gravity theory in 4D and theory of 3-forms
in 7D is to compare their actions. The action principle that entails the relation
$d{}^*\Omega=0$ as its Euler--Lagrange equation is the total volume of the space as
computed using the metric defined by $\Omega$, see \cite{Hitchin-67} and below. When one
computes this 7D functional on the ansatz (\ref{omega}), one finds a constant multiple of
the volume of the 4D base computed using (\ref{volume}). In other words, on ansatz
(\ref{omega}), the 7D action functional reduces to the action of the 4D theory of
connections. This relation between the action functionals makes it less surprising that
their critical points are related.

We now proceed to describing all constructions in more detail. We start by reviewing some
basic facts about 3-forms in 7 dimensions and their relation to $G_2$ holonomy. We then
describe in Section \ref{sec:gravity} the diffeomorphism-invariant ${\rm SO}(3)$ gauge
theory that gives rise to (\ref{det-eqs}). In Section \ref{sec:BS}, we review the
construction due to Bryant and Salamon. We present our generalisation of this
construction in Section \ref{sec:us}, and give examples of metrics that arise in this way
in Section \ref{sec:examples}. We conclude with a discussion.

\section{3-Forms in 7 dimensions and $G_2$-holonomy manifolds}

The material in this section is standard  (see, e.g., \cite{Hitchin-67}) and is reviewed
for the convenience of the reader. It was stunning for us to realise that the beautiful
geometry reviewed below has been known for more than a century, see \cite{History}.
In particular the characterisation of $G_2$ via 3-forms is a result due to Engel from 1900.

\subsection{Stable 3-forms}

Let us start with some linear algebra in $\R^7$. A 3-form $\Omega\in \Lambda^3 \R^7$ is called {\it stable} if it lies in a open orbit under the action of ${\rm GL}(7)$, see \cite{Hitchin-67}. This notion gives a generalisation of non-degeneracy of forms and implies that any nearby form can be reached by a ${\rm GL}(7)$ transformation. Thus, stable 3-forms can also be called generic or non-degenerate.

For real 3-forms, there are exactly two distinct open orbits, characterised by the sign of a certain invariant, see below, each of which is related to a real form of $G_2^\C$. In this paper we are mostly concerned with the open orbit corresponding to the compact real form $G_2$. For every such $\Omega$, there exists a set $\theta^1,\ldots, \theta^7$ of 1-forms 
in which $\Omega$ is expanded in the following canonical form:
\be\label{review-canonical}
\Omega = \theta^5\wedge \theta^6 \wedge \theta^7 + \theta^5 \wedge \Sigma^1
+ \theta^6\wedge \Sigma^2 + \theta^7 \wedge \Sigma^3,
\ee
where
\be\label{sigmas}
\Sigma^1 = \theta^1\wedge \theta^2 - \theta^3\wedge \theta^4, \quad \Sigma^2 = \theta^1\wedge \theta^3 - \theta^4\wedge \theta^2, \quad \Sigma^3 = \theta^1\wedge \theta^4 - \theta^2\wedge \theta^3.
\ee
Here the particular combinations $\Sigma^i$ are motivated in relation to (anti-)self-duality in 4 dimensions and are related to the embedding of ${\rm SO}(3)$ into ${\rm SO}(4)\subset G_2$. The relation to anti-self-dual 2-forms in 4 dimensions will be important in the construction below. 

The fact of central importance about stable 3-forms in 7 dimensions is that a stabilizer of such a
form in ${\rm GL}(7)$ is isomoprhic to the exceptional Lie group $G_2$. This group has
dimension $14$, and this number arises as the dimension 49 of ${\rm GL}(7)$ minus the
dimension 35 of $\Lambda^3 \R^7$. Thus, the space of stable 3-forms is the homogeneous
group manifold ${\rm GL}(7)/G_2$.

We can then generalise the notion of stable forms to 3-forms on a 7-dimensional differentiable manifold $M$. These are forms that are stable at every point.

\subsection{The metric}

The basic fact about stable 3-forms  on a 7-dimensional manifold $M$ is that they
naturally define a metric in $M$ by the relation
\be\label{3form-metric}
g_\Omega(\xi,\eta) v_{g_\Omega} = i_\xi \Omega \wedge i_\eta \Omega \wedge \Omega \, .
\ee
Here, $v_{g_\Omega}$ is the metric volume form, and $i_\xi$ denotes the operation of
insertion of a vector into a form.  The sign of the metric volume form $v_{g_\Omega}$ is
uniquely fixed by the requirement that the metric defined by \eqref{3form-metric} has
specific (say, Euclidean) signature.  In this way, a 3-form $\Omega$ defines both the
metric $g_\Omega$ and the orientation, corresponding to $v_{g_\Omega}$.

It is then a simple computation that, for a 3-form presented in the canonical form
(\ref{review-canonical}), the arising metric is
\be\label{review-metric}
g_\Omega = \sum_{I=1}^{7} \theta^I \otimes \theta^I \, ,
\ee
and the orientation is given by $\theta^1 \wedge \cdots \wedge \theta^7$. Given that
$G_2$ is the stabilizer of (\ref{review-canonical}), it also stabilizes metric
(\ref{review-metric}). This gives an embedding $G_2\subset {\rm SO}(7)$.

What we have reviewed above concerns the compact real form $G_2$ of $G_2^\C$. There is also the
orbit of real 3-forms that is related to the non-compact real form of $G_2^\C$. Such 3-forms
also have a canonical form similar to (\ref{review-canonical}), but with some signs
changed. In exactly the same way as \eqref{3form-metric}, they give rise to a metric of
signature $(3,4)$.

The counting of components shows that 3-forms contain more information than just that of
a metric. Indeed, to specify a metric in 7 dimensions, we need $7\times 8/2=28$ numbers,
while the dimension of the space of 3-forms is $35$. Thus, there are $7$ more components in a 3-form. 
These correspond to components of a unit spinor, see \cite{Witt} for more details.

\subsection{A functional}

Given a stable 3-form, we construct the metric and the corresponding volume form as
above.  We can integrate this volume form over the manifold to get the functional
\be\label{review-func}
S[\Omega] = \int_M v_{g_\Omega} \, .
\ee
This functional can also be computed explicitly, without computing the metric, via the
following construction. Let $\tilde{\epsilon}^{\alpha_1\ldots \alpha_7}$ be the canonical
anti-symmetric tensor density that exists independently of any metric. Then construct
\be\label{review-object}
\Omega_{\alpha_1\beta_1\gamma_1} \ldots \Omega_{\alpha_4\beta_4\gamma_4}
\tilde{\Omega}^{\alpha_1\ldots\alpha_4}\tilde{\Omega}^{\beta_1\ldots\beta_4}\tilde{\Omega}^{\gamma_1\ldots\gamma_4}
\, ,
\ee
where
\be
\tilde{\Omega}^{\alpha_1\ldots\alpha_4} := \tilde{\epsilon}^{\alpha_1\ldots\alpha_7}
\Omega_{\alpha_5\alpha_6\alpha_7}\, .
\ee
Then the object (\ref{review-object}) is of homogeneity degree $7$ in $\Omega$, and has
density weight $3$. Its cube root then has the right density to be integrated over the
manifold. The functional constructed in this way is a multiple of (\ref{review-func}). 

It is interesting to note 
that the invariant (\ref{review-object}) has been known already to Engel in 1900, see \cite{History}. This invariant gives 
a useful stability criterion: a form $\Omega$ is stable iff (\ref{review-object}) (equivalently (\ref{review-func})) is non-zero. The sign of this invariant then allows to distinguish between the two ${\rm GL}(7)$ 3-form orbits described above. 

\subsection{The first variation}

As explained in \cite{Hitchin-67}, the first variation of the functional
(\ref{review-func}) in $\Omega$ has a simple form
\be\label{var}
\delta S[\Omega] \sim \int_M {}^* \Omega \wedge \delta \Omega \, .
\ee
The precise numerical coefficient in this equation is of no importance for us. The 4-form
${}^*\Omega$ is just the Hodge dual of $\Omega$ computed with respect to the metric
defined by $\Omega$.

\subsection{Holonomy reduction}

The fundamental result due to Alfred Gray \cite{Gray} states: Let $\Omega\in \Lambda^3 M$
be a 3-form on a 7-manifold. Then $\Omega$ is parallel with respect to the Levi-Civita
connection of $g_\Omega$ iff $d\Omega=0$ and $d{}^*\Omega=0$. In other words, the
condition of $\Omega$ being parallel with respect to the metric it defines is equivalent
to the conditions of $\Omega$ being closed and co-closed, where co-closedness is again
with respect to the metric it defines.

The next basic fact is that if a Riemannian manifold $(M, g)$ has a parallel 3-form
$\Omega$, then the holonomy group of $M$ is contained in $G_2$. In this paper, we will
not be concerned whether the holonomy group is all of $G_2$ or is just contained in it,
and will simply refer to 7-manifolds $M$ with 3-forms satisfying $d\Omega=0$ and
$d{}^*\Omega=0$ as $G_2$-holonomy manifolds. Techniques for proving that the holonomy
equals $G_2$ can be found in \cite{BS}.

Combining Gray's result with the formula (\ref{var}) for the first variation of the
functional $S[\Omega]$, we see that $G_2$-holonomy manifolds are critical points of
$S[\Omega]$, provided one varies $\Omega$ in a fixed cohomology class $\delta \Omega =
dB$, $B\in \Lambda^2 M$. This variational characterisation is explored in depth in
\cite{Hitchin-67}.

\section{Diffeomorphism-invariant ${\rm SO}(3)$ gauge theories and gravity}
\label{sec:gravity}

In this section, we review how gravity theories in 4D (including General Relativity) can
be described as diffeomorphism-invariant theories of ${\rm SO}(3)$ connections. This
material is mainly from \cite{Krasnov:2011up, Krasnov:2011pp}, see also
\cite{Fine:2013qta} for a more mathematical exposition.

\subsection{Volume functionals from ${\rm SO}(3)$ connections}

As before, let $A^i$ be an ${\rm SO}(3)$ connection in an associated $\R^3$ bundle over a
4-dimensional manifold $M$, and let $F^i = dA^i + (1/2) \epsilon^{ijk} A^j\wedge A^k$ be
its curvature. Choose an orientation (volume) form $v$ on $M$, and define the matrix
$X^{ij}$ via
\be\label{X}
F^i \wedge F^j = - 2 X^{ij} v \, .
\ee
Since different orientation forms are related by multiplication by a nowhere vanishing
function, it is clear that $X^{ij}$ is defined only modulo such multiplication. The
factor of $2$ here is for future convenience. The choice of minus sign has to do with
later identification of $F^i$'s with anti-self-dual forms.

Let $f(X^{ij})$ be a function from symmetric $3\times 3$ matrices to reals satisfying the
following two requirements: (i)~it is gauge invariant $f(OXO^T)=f(X)$, where $O\in {\rm
SO}(3)$; (ii)~it is homogeneous of degree one, $f(\alpha X) = \alpha f(X)$ for any real
$\alpha$. It is clear that any such function can be applied to the wedge product of
curvatures
\be\label{f-FF}
f(F^i\wedge F^j) := - 2 f(X^{ij}) v \, ,
\ee
and that the result is a well-defined and gauge invariant 4-form on $M$. Thus, any such
function gives rise to a diffeomorphism and gauge invariant functional of connections
\be\label{action}
S_f[A] = - \frac{1}{2} \int_M f(F^i\wedge F^j) \, ,
\ee
where integration is performed with respect to the orientation $v$. Note that this
functional is just the total volume of $M$ computed using the volume form constructed
from the curvature of $A$. Thus, any choice of function $f$ gives rise to a
diffeomorphism-invariant theory of ${\rm SO}(3)$ connections.  

It is clear that there are many functions $f$ satisfying the required properties. An easy way to count is to diagonalise the matrix $X$. The function $f$ is then a homogeneity degree one function of the eigenvalues. There are as many such functions as functions of two variables. We will describe some most interesting choices of $f$ below.

\subsection{Euler--Lagrange equations}

The extrema of (\ref{action}) are connections satisfying the following second order PDE's
\be\label{feqs}
d_A \left( \frac{\partial f}{\partial X^{ij}} F^j \right) = 0 \, .
\ee
Note that the matrix of derivatives of the function $f$ with respect to $X$ is
homogeneity degree zero in $X$, and is hence well-defined even though $X$ is only defined
modulo multiplication by a function.  In other words, equations \eqref{feqs} are
independent of the choice of the orientation form $v$ in the definition \eqref{X}
of $X$.

\subsection{Definite connections and the choice of orientation}

A preferred orientation of $M$ can be fixed in the case of an important class of {\it
definite\/} connections, see \cite{Fine:2013qta}. A connection $A^i$ is called definite
if the corresponding matrix $X^{ij}$ defined via (\ref{X}) is definite, i.e., all its
eigenvalues are of the same sign. Then a preferred orientation of $M$ is represented by
an orientation form $v$ for which the matrix $X$ is positive definite.

In what follows, we will always use the orientation provided by the connection. In particular, we use the orientation that makes $X$ a positive definite matrix in defining the action (\ref{action}). Note that this does not
mean that the functional (\ref{action}) is always positive definite. For example, the
function $f(X) = - {\rm Tr\,} X$ gives a negatively oriented volume
form. In this paper, in order to avoid confusion, we will always use functions $f$ that give volume forms of the same orientation as is provided by the connection. Thus, our action functionals here will always be of one (positive) sign.\footnote{\label{foot}While the overall sign of the action is not important in the
pure gravity theory, its sign relative to the action of other fields will, of course, be
important.}

\subsection{Metrics from definite connections}

An ${\rm SO}(3)$ connection that satisfies a rather weak requirement that it is definite
defines a conformal structure of a Riemannian metric on $M$. This is the conformal class
already defined in \eqref{urbanc}.  This is often referred to (especially in the physics
literature) as the Urbantke metric, as it was first introduced in \cite{Urbantke}. The
significance of this (conformal) metric is that it is the unique conformal structure with
respect to which the triple of curvature 2-forms is anti-self-dual.  

To complete the definition of the metric we need to specify the volume form. As is explained above, any choice of function $f$ (satisfying gauge invariance and homogeneity properties) gives a volume form. Thus, any choice of $f$ defines a metric in the conformal class of (\ref{urbanc}).

Thus, once a choice of $f$ is made, we have a metric defined by the connection. When the connection satisfies its Euler-Lagrange equations (\ref{feqs}), the metric defined by $A$ is constrained. Below we shall see that Einstein metrics can be obtained in exactly this way, for a certain choice of $f$.

\subsection{The natural choice}
\label{sec:natural}

Even though there exists freedom in choosing the conformal factor in (\ref{urbanc}), there exists a mathematically natural choice. We shall refer to the mathematically natural choice of the metric as {\it the Urbantke metric} $g_{\rm U}$.

The connection provides an orientation (in which $X$ is positive definite), and we choose the metric volume form
to be positively oriented. We also require the metric to be of Riemannian (all plus) signature. The Urbantke metric is then defined via 
\be\label{Urb*}
g_{\rm U} (\xi,\eta) v_{\rm U} = \frac{\sigma}{6} \epsilon^{ijk} i_\xi F^i \wedge i_\eta
F^j \wedge F^k \, ,
\ee
where $v_{\rm U}$ is the metric volume form, and where $\sigma = \pm 1$ is the sign that
depends on the connection.  This sign $\sigma$ in \eqref{Urb*} is called {\em the sign of
the definite connection\/} $A^i$.  It is discussed in more detail in Section~2.2 of
\cite{Fine:2013qta}, see also below. This sign is necessary in (\ref{Urb*}) to give the Urbantke metric the all plus signature.

\subsection{A computation}

As we know from above, any volume form constructed from the curvature corresponds to some
choice of $f$. Let us see what this choice is for the Urbantke metric (\ref{Urb*}). 

As we already mentioned above, any metric in the conformal class of (\ref{urbanc}) makes
the triple of curvature 2-forms anti-self-dual (ASD). Let us choose {\it some} metric $g$ in this conformal class, and 
introduce a canonical orthonormal basis $\Sigma^i$ in the space of ASD 2-forms for the metric $g$. Explicitly, given a frame basis, $\Sigma^i$'s are the forms that are given by (\ref{sigmas}). They satisfy the following algebraic relations
\be\label{SS}
\frac{1}{2} \Sigma^i\wedge \Sigma^j = - \delta^{ij} v_g \, ,
\ee
\be\label{SS-alg}
\Sigma^i_{\mu}{}^\rho \Sigma^j_{\rho}{}^\nu = \epsilon^{ijk} \Sigma^k_{\mu}{}^\nu -
\delta^{ij} \delta_{\mu}{}^{\nu} \, ,
\ee
where the space indices are raised by the metric inverse of $g$. The minus sign
in \eqref{SS} has to do with our usage of ASD forms rather than SD ones. The volume form $v_g$ in (\ref{SS})
is the metric volume form, positively oriented in the orientation provided by the connection. 

Then the curvature 2-forms can be expanded in the basis of $\Sigma^i$ as
\be\label{FS}
F^i = \sigma \left( \sqrt{X} \right)^{ij} \Sigma^j \, ,
\ee
where $\sigma = \pm 1$ is the sign of the definite connection $A^i$ already introduced in the
previous subsection, and $\sqrt{X}$ is the positive-definite matrix square root of the
positive-definite matrix $X$. We stress that the relation (\ref{FS}) can be written for an arbitrary choice of metric $g$ in the 
conformal class of (\ref{urbanc}). This relation can also be used as an alternative definition of the sign of the definite connection. The decomposition (\ref{FS}) follows using
(\ref{SS}). Indeed, we have $F^i\wedge F^j = \sigma^2 \sqrt{X}^{ik} \sqrt{X}^{jl} (-2)
\delta^{kl} v_{g}= - 2X^{ij} v_{g}$. 

We now use (\ref{FS}) with $\Sigma^i$'s being those for the Urbantke metric (\ref{Urb*}). Thus, we now take $X = X_{\rm U}$ with respect to the volume form of the metric $g_{\rm U}$. Substituting \eqref{FS} into (\ref{Urb*}) 
and using \eqref{SS-alg}, we get the
relation $g_{\rm U} = \left( \det X_{\rm U} \right)^{1/2} g_{\rm U}$, from which we
conclude that
\be\label{det-cond}
\det X_{\rm U} = 1\, .
\ee

As we already remarked, the sign (and even the overall factor) of the Lagrangian function $f (X)$ in action
\eqref{action} does not matter in the pure-gravity theory (see, however, footnote
\ref{foot} on page \pageref{foot}). We can then always take this function to be
positive-valued for positive-definite $X$. We then note that for any function $f$ we can use the
volume form $v_f = f(X) v$ to define $X$. One then has $v_f = f(X_f) v_f$ and hence $f(X_f)=1$. This immediately allows us to translate the condition (\ref{det-cond}) into a choice of the function $f$. Thus, the
condition (\ref{det-cond}) derived above corresponds to a homogeneous degree one function
\be\label{f*}
f (X) = \left( \det X \right)^{1/3}\, .
\ee

We then note that for this function
\be
\frac{\partial f}{\partial X} = \frac13 \left( \det X \right)^{1/3} X^{-1}\, ,
\ee
and so the field equations (\ref{feqs}) reduce to (\ref{det-eqs}). 
As clear from the preceding subsection, a characteristic property of this function is
that $f \left( F \wedge F \right)$ coincides with the volume form of the Urbantke metric
$g_{\rm U}$ defined in \eqref{Urb*}.

\subsection{Einstein connections}

There is a {\it different\/} choice of $f (X)$ that gives rise to Einstein metrics
\cite{Krasnov:2011pp}. Let us define
\be\label{f-GR}
f_{\rm GR} = \left( {\rm Tr\,} \sqrt{X} \right)^2 \, .
\ee
We then have
\be
\frac{\partial f_{\rm GR}}{\partial X} = \left( {\rm Tr\,} \sqrt{X} \right) X^{-1/2} \, ,
\ee
and
\be
\frac{\partial f_{\rm GR}}{\partial X^{ij}} F^j =  \sigma \, {\rm Tr\, } \sqrt{X}  \, \Sigma^i \, ,
\ee
where we have used (\ref{FS}). This is valid for $X$ and $\Sigma^i$'s defined as in (\ref{FS}) with respect
to some metric in the conformal class of (\ref{urbanc}). 

We can then fix the metric $g_{\rm GR}$ in the conformal class of (\ref{urbanc}) so that 
\be
{\rm Tr\,} \sqrt{X_{\rm GR}} = 1 \, .
\ee
Once the metric is fixed in this way, the field equations (\ref{feqs}) become $d_A \Sigma^i_{\rm GR}=0$, where $\Sigma^i_{\rm GR}$ is the basis (\ref{sigmas}) of ASD forms for the metric $g_{\rm GR}$. This equation is equivalent to the
statement that the connection $A$ is the anti-self-dual part of the Levi-Civita
connection for the metric with the basis of ASD 2-forms $\Sigma^i_{\rm GR}$. We then have a metric
with the curvature of the ASD part of the Levi-Civita connection being ASD as a 2-form.
This is known to be equivalent to the Einstein condition. The arising metrics are
Einstein with the cosmological constant $\Lambda = 3\sigma$.

For more information about General Relativity in the language of connections the reader is
referred to exposition in \cite{Fine:2013qta}. The choice of $f (X)$ that leads to GR is
not to play any further role in this paper, and is described here just to illustrate the
statement that it is the mathematically more natural choice (\ref{f*}) that plays role in the
construction of the $G_2$ holonomy metrics, not (\ref{f-GR}).

Another way to state that a pure-connection theory with Lagrangian (\ref{f*}) is not GR
is to say that connections satisfying (\ref{det-eqs}) give rise to metrics (\ref{Urb*})
that are not Einstein. It would be interesting to characterise the arising metrics is
some way. 

\subsection{Generality of the volume functionals}

Even though this has little to do with the main subject of this paper, let us remark that
the parametrisation (\ref{FS}) of the curvature makes it clear that the only
gauge-invariant volume form that can be constructed from the curvature of the connection
is of the type (\ref{f-FF}) for some function $f$. This follows from the fact that the
volume form can only be constructed from factors of the curvature and the anti-symmetric
tensor $\tilde{\epsilon}^{\mu\nu\rho\sigma}$  that has density weight one and that exists
on any manifold. Using (\ref{FS}) to parametrise the curvature, as well as the fact that
$\Sigma$'s are anti-self-dual, one can convince oneself that all contractions of the
spacetime indices are taken care of by the algebra (\ref{SS-alg}), and that what remains
is some gauge-invariant scalar built from factors of the matrix $\sqrt{X}$. Thus, for
${\rm SO}(3)$ connections that define a conformal class of metrics, all gauge-invariant
functions of the curvature are of the type (\ref{f-FF}). One can easily generalise the
construction (\ref{f-FF}) to other gauge groups, but in that case there are functions of
the curvature that do not reduce to (\ref{f-FF}). One should keep in mind this special
character of the ${\rm SO}(3)$ theory.

\subsection{Instanton solutions}

For any function $f (X)$, connections satisfying $X^{ij}\sim \delta^{ij}$ give rise to
metrics that are self-dual Einstein. Indeed, in this case the field equations
(\ref{feqs}) reduce to the Bianchi identity for the curvature and are automatically
satisfied for any $f (X)$. When $X^{ij}\sim \delta^{ij}$, there exists a metric with
respect to whose volume form $F^i = \sigma \Sigma^i$. The Bianchi identity then states
that $d_A \Sigma^i=0$ and, therefore, $A$ is the ASD part of the Levi-Civita connection.
It is clear that the corresponding metric is Einstein, as there is no SD part in the
curvature 2-form of $A$. Also, because $F^i=\sigma \Sigma^i$, the ASD part of the Weyl
tensor vanishes (${\rm Weyl}_- = 0$), and we have a self-dual Einstein metric of scalar
curvature $12\sigma$.

Thus, self-dual Einstein metrics corresponding to connections with $X^{ij}\sim
\delta^{ij}$ are solutions of (\ref{feqs}) for any $f (X)$. In particular, these
solutions are shared by theory (\ref{f*}) and GR (\ref{f-GR}).

\subsection{More general solutions}

Even though we are far from understanding all Einstein metrics on 4-manifolds, some
intuition as to how many solutions there exist comes from the Lorentzian version of the
theory. Indeed, GR with Lorentzian signature is a theory with local degrees of freedom,
and so the space of solutions is infinite-dimensional. For example, solutions can be
obtained by evolving the initial data.

A similar description is also possible in the Riemannian context, in particular in the
setting of asymptotically hyperbolic metrics. Then, as is well known from the work
\cite{FG}, one can solve for asymptotically hyperbolic Einstein metrics in the form of an
expansion in powers of the `radial' coordinate. The free data for this expansion are a
conformal class of metric on the boundary (modulo boundary diffeomorphisms), together
with a symmetric traceless transverse tensor that appears as free data in some higher
order of the expansion. There are $2+2$ free functions on the boundary as free data, and
this is the Riemannian analog of the statement that GR has 2 propagating degrees of
freedom.

A similar expansion in the language of connections was developed in
\cite{Fine:2015hef}. One outcome of the analysis of this paper is that the expansion
is universal for the whole class of theories (\ref{action}), whatever the function $f$
is. Only the details of the expansion at sufficiently high order in the radial coordinate
start to depend on $f$. In the first few orders, the expansion is completely independent
of $f$. In particular, the count of free data that seeds the expansion is
$f$-independent. This means that the free data to be prescribed to get an asymptotically hyperbolic solution of 
theory (\ref{f*}) (locally near the boundary) are $2+2$ free functions on the 3-dimensional
asymptotic boundary. This illustrates the
statement that the theory (\ref{f*}) has as many solutions as GR.

Some explicit cohomogeneity one asymptotically hyperbolic solutions of theory (\ref{f*}) are described below. 

\section{Bryant--Salamon construction}
\label{sec:BS}

We now review the construction of \cite{BS}  using the notation compatible with our
discussion of the ${\rm SO}(3)$ formulation of gravity.

\subsection{Ansatz}

Let $(M,g)$ be a self-dual Einstein 4-manifold, and let $\Sigma^i$, $i=1,2,3$, be the
basis of ASD 2-forms satisfying properties (\ref{SS}) and (\ref{SS-alg}). For example,
the 2-forms $\Sigma^i$ can be constructed from the frame 1-forms via (\ref{sigmas}). Let
$A^i$ be the ASD part of the Levi-Civita connection. This is an ${\rm SO}(3)$ connection
that satisfies
\be
d_A \Sigma^i =0\, .
\ee
The self-dual Einstein condition translates into
\be\label{inst}
F^i = \sigma \Sigma^i \, , \qquad \sigma=\pm 1\, ,
\ee
where we have normalised our metric so that the scalar curvature is $12\sigma$.

An arbitrary ASD 2-form can be written as $\Sigma(y) = \Sigma^i y^i$, and so the
quantities $y^i$ are the fiber coordinates in the bundle of ASD 2-forms over $M$. We make
the following ansatz for the calibrating 3-form:
\be\label{BS}
\Omega = \frac{1}{6} \alpha^3\epsilon^{ijk} d_A y^i \wedge d_A y^j \wedge d_A y^k  +
2\alpha\beta^2  d_A y^i \wedge \Sigma^i\, ,
\ee
where $d_A y^i = dy^i + \epsilon^{ijk} A^j y^k$ is the covariant derivative with respect
to $A$, and $\alpha$ and $\beta$ are functions of $y^2$.

\subsection{Closure}

We now require the form $\Omega$ to be closed. Because $\Omega$ does not have any
internal indices we can apply the covariant derivative instead of the exterior one. When
differentiating the first term, we only need to differentiate the quantities $d_A y^i$,
as differentiating $\alpha$ would lead to exterior products of four one-forms from the
triple $\{ d_A y^i \}$, which are zero. In the second term, we do not need to apply the
derivative to $\Sigma^i$ because it is covariantly closed. We also do not need to
differentiate $d_A y^i$ since this produces a multiple of $\epsilon^{ijk} F^j y^k \wedge
\Sigma^i$, which is equal to zero due to (\ref{inst}) and (\ref{SS}). We thus get
\be
d\Omega = \frac{1}{2} \alpha^3 \epsilon^{ijk} \epsilon^{ilm} F^l y^m \wedge d_A y^j
\wedge d_A y^k + 2 \left( \alpha\beta^2 \right)' \left( 2y^i d_A y^i \right) \wedge
\left( d_A y^j \wedge \Sigma^j \right)\, .
\ee
We now use (\ref{inst}) and decompose the product of two epsilon tensors into products of
Kronecker deltas. We obtain
\be
d\Omega = \left[ - \sigma \alpha^3 + 4 \left( \alpha\beta^2 \right)' \right] \left( y^i
d_A y^i \right) \wedge \left( d_A y^i \wedge \Sigma^j \right) \, .
\ee
Thus, we must have
\be\label{BS-eqn-1}
4 \left( \alpha\beta^2 \right)' = \sigma \alpha^3
\ee
in order for the form to be closed. The quantity $\sigma=\pm 1$ is the sign already encountered above, see (\ref{FS}). 

\subsection{Canonical form}

We now compute the metric defined by $\Omega$, as well as its Hodge dual. The easiest way
to do this is to write the 3-form in the canonical form, so that the metric and the dual
form are immediately written. Thus, let $\theta^1, \ldots, \theta^7$ be a set of 1-forms
such that the 3-form $\Omega$ is
\ber\label{canonical}
\Omega &=& \theta^5\wedge \theta^6 \wedge \theta^7 + \theta^5 \wedge \left(
\theta^1\wedge \theta^2 - \theta^3\wedge \theta^4 \right) \nonumber \\ &&{}+
\theta^6\wedge \left( \theta^1\wedge \theta^3 - \theta^4\wedge \theta^2 \right) +
\theta^7 \wedge \left( \theta^1\wedge \theta^4 - \theta^2\wedge \theta^3 \right) \, .
\eer
Then the 1-forms $\theta$ are an orthonormal frame for the metric determined by $\Omega$
\be
g_\Omega = \left( \theta^1 \right)^2 + \ldots + \left( \theta^7 \right)^2\, ,
\ee
and the Hodge dual ${}^*\Omega$ of $\Omega$ is given by
\ber
{}^*\Omega &=& \theta^1\wedge \theta^2\wedge \theta^3\wedge \theta^4 - \left(
\theta^1 \wedge \theta^2 - \theta^3\wedge \theta^4 \right) \wedge \theta^6\wedge \theta^7 \nonumber \\
&&{}- \left( \theta^1\wedge \theta^3 - \theta^4\wedge \theta^2 \right) \wedge
\theta^7\wedge \theta^5 - \left( \theta^1\wedge \theta^4 - \theta^2\wedge \theta^3
\right) \wedge \theta^5\wedge \theta^6\, .
\eer

\subsection{Calculation of the metric and the dual form}

We now put ansatz (\ref{BS}) into the canonical form (\ref{canonical}), and compute the
associated metric and the dual form. The canonical frame is easily seen to be
\be
\theta^{4+i} = \alpha d_A y^i, \qquad \theta^I = \beta \sqrt{2} e^I, \qquad I=1,2,3,4\, ,
\ee
where $e^I$ is the orthonormal frame such that the basis of ASD 2-forms is given by (\ref{sigmas}). 
The metric is then
\be
g_\Omega = \alpha^2 \sum_i \left( d_A y^i \right)^2 + 2\beta^2 \sum_I \left( e^I
\right)^2\, ,
\ee
and the dual form is
\be\label{BS*}
{}^*\Omega = -\frac{2}{3} \beta^4 \Sigma^i \wedge \Sigma^i - \beta^2 \alpha^2
\epsilon^{ijk} \Sigma^i \wedge d_A y^j \wedge d_A y^k \, .
\ee

\subsection{Co-closure}

We now demand the 4-form (\ref{BS*}) also to be closed. The first point to note is that
when we apply the covariant derivative to the factor $\beta^2 \alpha^2$ in the second
term, we generate a 5-form proportional to the volume form of the fiber. There is no such
term arising anywhere else, and we must demand
\be\label{BS-eqn-2}
\alpha\beta = {\rm const }
\ee
in order for (\ref{BS*}) to be closed. Differentiation of the rest of the terms gives
\be
d{}^*\Omega = -\frac{2}{3} \left( \beta^4 \right)' \left( 2y^i d_A y^i \right) \wedge
\Sigma^j\wedge \Sigma^j - 2 \beta^2\alpha^2 \epsilon^{ijk} \Sigma^i \wedge \epsilon^{jlm}
F^l y^m \wedge d_A y^k \, .
\ee
We now use (\ref{inst}) and (\ref{SS}) to get
\be
d{}^*\Omega = -\frac{2}{3} \left[ \left( \beta^4 \right)' - \sigma\beta^2\alpha^2 \right]
\left( 2y^i d_A y^i \right) \wedge \Sigma^j \wedge \Sigma^j \, ,
\ee
and so we must have
\be\label{BS-eqn-3}
\left( \beta^4 \right)' = \sigma\beta^2\alpha^2\, .
\ee

\subsection{Determining $\alpha$ and $\beta$}

The overdetermined system of equations (\ref{BS-eqn-1}), (\ref{BS-eqn-2}) and
(\ref{BS-eqn-3}) is nevertheless compatible. Without loss of generality, we can simplify
things and rescale $y^i$ (and therefore $\alpha$) so that
\be
\alpha\beta =1 \, .
\ee
With this choice, we have only one remaining equation to solve, which gives
\be\nonumber
\beta^4 = k + \sigma y^2,
\ee
where $k$ is an integration constant. We can then further rescale $y$ and $\beta$,
keeping $\alpha\beta=1$, to set $k = \pm 1$ at the expense of multiplying the 3-form
$\Omega$ by a constant. After all these rescalings, we get the following incomplete
solutions:
\be \begin{array}{ll}
\sigma = 1\, , \quad &\beta = (y^2 - 1)^{1/4}\, , \quad y^2>1 \, , \\
\sigma = -1\, , \quad &\beta = (1-y^2)^{1/4}\, , \quad y^2<1\, ,
\end{array}
\ee
as well as a complete solution for the positive scalar curvature:
\be
\sigma = 1, \quad \beta = (1+y^2)^{1/4} \, .
\ee
The two most interesting solutions, the incomplete solution for $\sigma=-1$ and the
complete solution for $\sigma=+1$, can be combined together as
\be
\beta = (1+\sigma y^2)^{1/4} \, .
\ee

\section{Construction in Theorem \ref{theo}}
\label{sec:us}

We now give details of our generalisation of the Bryant--Salamon construction.

\subsection{Ansatz and closure}

We parametrise the 3-form by an ${\rm SO}(3)$ connection in an $\R^3$ bundle over $M$:
\be\label{our-3}
\Omega = \frac{1}{6} \alpha^3\epsilon^{ijk} d_A y^i \wedge d_A y^j \wedge d_A y^k  +
2\sigma \alpha\beta^2  d_A y^i \wedge F^i\, ,
\ee
where the factor $\sigma=\pm 1$ is the sign of the definite connection. It is introduced
in the ansatz so that (\ref{our-3})  reduces to (\ref{BS}) for instantons (\ref{inst}).
It is then easy to see, using the Bianchi identity $d_A F^i=0$, that the condition of
closure of (\ref{our-3}) is unmodified and is still given by (\ref{BS-eqn-1}).

\subsection{The canonical form and the metric}

We now put (\ref{our-3}) into the canonical form (\ref{canonical}). To this end, we use
the parametrisation (\ref{FS}) of the curvature. It is then clear that the 1-forms
$\theta^{4+i}$ are some multiples of $\alpha \sqrt{X}^{ij} d_A y^j$. The correct factors
are easily found, and we have
\be
\theta^{4+i} = \left( \det X \right)^{-1/6} \alpha \left( \sqrt{X} \right)^{ij} d_A y^j,
\qquad \theta^I = \beta \sqrt{2} \left( \det X \right)^{1/12} e^I,
\ee
where $e^I$ are the frame 1-forms for the metric which makes $F^i$ anti-self-dual and
whose volume form is used to define the matrix $X^{ij}$, see (\ref{X}).

The metric determined by (\ref{our-3}) is then
\be\label{metric-om}
g_\Omega = \alpha^2  \left( \det X \right)^{-1/3} d_A y^i X^{ij} d_A y^j + 2\beta^2
\left( \det X \right)^{1/6} \sum_I \left( e^I \right)^2\, .
\ee

\subsection{The dual form and the co-closure}

The dual form reads
\be
{}^*\Omega = -\frac{2}{3} \beta^4 \left( \det X \right)^{1/3} \left( X^{-1} F \right)^i
\wedge F^i - \sigma \beta^2 \alpha^2  \left( \det  X \right)^{1/3} \left( X^{-1} F
\right)^i \epsilon^{ijk} \wedge d_A y^j \wedge d_A y^k\, ,
\ee
where again we expressed all ASD 2-forms on the base in terms of the curvature 2-forms
using (\ref{FS}). Note that, in both terms, the curvature appears either as itself, or in
the combination $\left( \det X \right)^{1/3} \left( X^{-1} F \right)^i$. It is now easy
to see that the same steps we followed in the Bryant--Salamon case can be repeated
provided
\be
d_A \left[ \left( \det X \right)^{1/3} X^{-1} F \right] = 0\, .
\ee
This is the field equation for theory (\ref{f*}) already quoted in (\ref{det-eqs}). The
Theorem stated in the Introduction is proven.

\subsection{Complete indefinite $G_2$ metrics for $\sigma=-1$}

We can modify our construction by not putting the sign $\sigma$ in front of the second
term in (\ref{our-3}). Then all of the construction goes unchanged except that $\sigma$
does not appear either in $\Omega$ or in ${}^*\Omega$. The differential equations for
$\alpha$ and $\beta$ then give $\beta=(1+y^2)^{1/4}$, and the metric is then complete in
the fiber direction for either sign. But the price one pays in this case is that the
second term in (\ref{metric-om}) will appear with a minus sign in front for $\sigma=-1$.
This will give a complete (in the fiber direction) metric of $G_2$ holonomy, but of
signature $(3,4)$ rather than a Riemannian metric.

\subsection{Metric induced on the base}

The 3-form (\ref{our-3}) defines the metric (\ref{metric-om}) on the total space of the
bundle. The metric induced on the base is in the conformal class that makes the curvature
2-forms $F^i$ anti-self-dual. The conformal factor can be read off from
(\ref{metric-om}). In particular, the corresponding volume form is
\be
v_\Omega = 4 \left( 1 + \sigma y^2 \right) \left( \det X \right)^{1/3} \epsilon \, ,
\ee
where $\epsilon$ is the orientation form used to define the matrix $X$. Thus, for a
constant $y^2$ the induced metric is a multiple of the metric that we already encountered
in the context of diffeomorphism-invariant ${\rm SO}(3)$ gauge theory defined by the
function (\ref{f*}).

We now remark that, in the context of ${\rm SO}(3)$ gauge theory, the metric
interpretation is possible, but nothing forces us to introduce this metric, as the theory
itself is about gauge fields, and metric is a secondary object. However, after embedding
into 7D, we see that the connection is a field that parametrises the closed 3-form
$\Omega$, and the 3-form naturally defines a metric in the total space of the bundle. In
the context of 7D theory, the metric arises more naturally and unavoidably. Since this 7D
metric induces a metric on the base, the 7D construction provides an explanation why the
metric should also be considered in the context of 4D ${\rm SO}(3)$ gauge theory.

\subsection{Relation between the 7D and 4D action functionals}

As is well-known (see \cite{Hitchin-67}), the co-closure condition $d{}^*\Omega$ can be
obtained by minimising a certain volume functional of $\Omega$ with respect to variations
of $\Omega$ by an exact form. The functional in question is just the volume of the 7D
manifold computed using the metric defined by $\Omega$. For our ansatz (\ref{our-3}), the
metric is given by (\ref{metric-om}). The fiber part gives the volume element $\alpha^3
(dy)^3$, while the base part gives $4\beta^4 \left( \det  X \right)^{1/3} \epsilon$.
Thus, the volume functional reduces for our ansatz to
\be
S[\Omega] = 4 \int d^3y  \left( 1 + \sigma y^2 \right)^{1/4} \int_M \left( \det X
\right)^{1/3} \epsilon \, .
\ee
This is proportional to the action of the ${\rm SO}(3)$ gauge theory on the base. In the
incomplete case $\sigma=-1$, the integral over the fiber (from $y=0$ to $y=1$) can be
taken and is finite, and we get
\be\label{relation}
S_{\sigma=-1}[\Omega] = \frac{16 \sqrt{\pi}\, \Gamma^2(1/4)}{21\sqrt{2}} \int_M \left(
\det X \right)^{1/3} \epsilon \, .
\ee
In either case, the volume functional for the 3-form (\ref{our-3}) in 7 dimensions is a
multiple of the volume functional for the ${\rm SO}(3)$ connection in 4D\@. Thus, there
is a relation not only between solutions of the two theories, but also between the action
functionals.

Let us note that we can also get relation (\ref{relation}) to work in the case
$\sigma=+1$ at the expense of making the 7D metric indefinite of signature $(3,4)$. This
is achieved just by putting the minus sign in front of the second term in (\ref{our-3})
also for the $\sigma=+1$ case. The 7D metric is then indefinite, but induces a Riemannian
signature metric on the base. In this case, the function $\beta = \left( 1 - y^2
\right)^{1/4}$, and so we get an incomplete metric in the fiber direction, and a finite
multiple relation (\ref{relation}) between the volumes.

\section{Examples}
\label{sec:examples}

We now describe some examples of 7D metrics obtained from the above construction. We
build 7D metrics from cohomogeneity one metrics on the base. We describe two easy
examples in which the base metrics are asymptotically hyperbolic. In our first example,
the asymptotic metric on the conformal boundary is that of $\R^3$, while, in the second
example, it is $S^1\times S^2$. Finally, we attempt the Bianchi IX case, but find
ourselves unable to solve the arising ODE's even in the bi-axial case. We still describe
this example, as it is likely the most interesting one from the mathematical viewpoint.

\subsection{Bianchi I}

The simplest, but still non-trivial example to consider is that of co-homogeneity one on
the base, with the base manifold having the structure $\R^3\times \R$\@. Thus, let
$dx^{1,2,3}$ be the one-forms in the $\R^3$ directions, and let $r$ be the coordinate in
the remaining direction, which we call `radial.' We make the following ansatz of the
connection:
\be
A^1 = a_1(r) dx^1, \qquad {\rm etc}\, .
\ee
The curvature forms read
\be
F^1 = a_1' dr\wedge dx^1 + a_2 a_3 \, dx^2\wedge dx^3, \qquad {\rm etc}\, .
\ee
We get the Euler--Lagrange equations by first evaluating the action of the full theory on
this ansatz, and then performing the variation.  Due to the symmetry of the problem, one
gets the same equations as would follow by substituting the ansatz into (\ref{det-eqs}).
The volume action evaluated on the above ansatz is then a multiple of
\be \label{biai}
S\sim \int dr \left( a_1' a_2' a_3' a_1^2 a_2^2 a_3^2 \right)^{1/3} \, .
\ee
The arising Euler--Lagrange equations read
\be
- \left( \frac{L}{a_1'} \right)' + \frac{2L}{a_1} = 0, \qquad {\rm etc} \, ,
\ee
where $L$ is the Lagrangian in \eqref{biai}. They can easily be solved by choosing the
radial coordinate so that
\be
L\equiv \left( a_1' a_2' a_3' a_1^2 a_2^2 a_3^2 \right)^{1/3} = 1 \, .
\ee
This gives
\be
a_1 = c_1 (r-r_1)^{1/3}, \qquad {\rm etc}\, ,
\ee
where $c_i$ and $r_i$ are integration constants.

The corresponding matrix $X$, calculated with respect to the orientation form $\epsilon =
dr\wedge d^3 x$, is given by
\be
X = \frac{1}{3} c_1 c_2 c_3 \left[ (r-r_1)(r-r_2)(r-r_3)\right]^{1/3} {\rm diag\,}\left(
\frac{1}{r-r_1},\frac{1}{r-r_2},\frac{1}{r-r_3}\right) \, .
\ee
We want our connection to be definite. Let us now assume that $c_1 c_2 c_3=3$, which is
always possible for $c_1 c_2 c_3>0$ by rescaling the metric. We can always order the
integration constants so that $r_3 < r_2 < r_1$. Then we get a definite connection for
$r> r_1$ and $r<r_3$.

Under these assumptions, we can easily compute the associated basis of ASD 2-forms
$\Sigma = \sigma X^{-1/2} F$. In order for the basis of $\Sigma$'s to take the canonical
form in the orientation of $dr\wedge d^3 x$, we need to take
\be
\sigma=-1 \, .
\ee
We get
\be
\Sigma^1 = - \left[ (r-r_2)(r-r_2) \right])^{-1/6} \left[ \frac{c_1 dr\wedge
dx^1}{3(r-r_1)^{1/3}} + c_2 c_3 \left( \prod_i (r-r_i)\right)^{1/3}  dx^2 \wedge
dx^3\right] \, ,
\ee
etc.  We can then easily determine the orthonormal basis for the metric:
\be
\theta^r = - \frac{1}{3}  \left[ \prod_i (r-r_i)\right]^{-1/3} dr, \qquad \theta^1 =c_1
\left[ (r-r_2)(r-r_2) \right]^{1/6} dx^1 \, , \qquad {\rm etc} \, ,
\ee
so that the metric is
\be\nonumber
ds^2_4 = \frac{1}{9} \left[ \prod_i (r-r_i)\right]^{-2/3} dr^2 +\left[ \prod_i
(r-r_i)\right]^{1/3} \left[  \frac{c_1^2  (dx^1)^2}{(r-r_1)^{1/3}} + \frac{c_2^2 (dx^2)^2
}{(r-r_2)^{1/3}}  +\frac{c_3^2 (dx^3)^2}{(r-r_3)^{1/3}} \right]\, .
\ee
For large $r$, the matrix $X$ approaches the identity matrix, so we have an
asymptotically self-dual Einstein solution. We can introduce the new coordinate
$r^{1/3}=\exp\rho$, in terms of which the asymptotic metric takes the form $ds^2 =
d\rho^2 + e^{2\rho}\sum_i c_i^2 (dx^i)^2$. This is the metric of the 4D hyperbolic space
with $\Lambda=-3$. The 7D lift of the full metric is given by (\ref{metric-om}):
\ber \label{bianchi-lift}
ds^2_7 &=& \frac{\left[ (r-r_1)(r-r_2)(r-r_3) \right]^{1/3}}{(1-y^2)^{1/2}}\sum_i
\frac{1}{r-r_i} \left[dy^i + \sum_{j,k} \epsilon^{ijk} c_j (r-r_j)^{1/3} dx^j
y^k\right]^2
\nonumber \\
&&{} + 2 \left( 1 - y^2 \right)^{1/2} ds_4^2 \, .
\eer

\subsection{Spherically symmetric solution}

We take the following spherically symmetric ansatz:
\be\label{conn-sph}
A^1 = a(R) dt + \cos \theta\,  d\phi\, , \qquad A^2 = -b (R) \sin \theta\,  d\phi\, ,
\qquad A^3 = b(R) d\theta\, ,
\ee
where $R$ is some radial coordinate. The curvatures are
\beq \label{curv-ss} \begin{array}{rcl}
F^1 &= &- a' dt\wedge dR + \left( b^2 - 1 \right) \sin \theta\,  d\theta \wedge d\phi\, , \medskip \\
F^2 &= &ab \, d\theta \wedge dt - b' \sin \theta\,  dR \wedge d\phi\, , \medskip \\
F^3 &= &- ab \sin \theta\,  dt \wedge d\phi + b' dR\wedge d\theta \, .
\end{array}
\eeq
The action evaluated on the ansatz reads
\be
S \sim \int dR \left[ a'\left( b^2 - 1 \right) a^2 \left( \left( b^2 - 1 \right)'
\right)^2 \right]^{1/3}\, .
\ee
Minimizing it with respect to $a$ and $b^2 - 1$, and again choosing the radial coordinate
in which $L=1$, we get
\be
- \left( \frac{1}{a'} \right)' + \frac{2}{a} =0 \, , \qquad - \left[ \frac{2}{\left( b^2
- 1 \right)'} \right]' + \frac{1}{b^2-1} = 0\, .
\ee
This integrates to
\be\label{ab}
a=C_1 \left( R - R_1 \right)^{1/3}\, , \qquad b^2 - 1 = C_2 \left( R - R_2
\right)^{2/3}\, ,
\ee
where $C_{1,2}$ and $R_{1,2}$ are integration constants. The corresponding matrix
$X^{ij}$, determined with respect to the orientation form $\epsilon = -dt \wedge dR
\wedge \sin \theta\, d \theta \wedge d\phi$, is
\be
X=\frac{C_1C_2}{3} {\rm diag}\left[ \left(\frac{R-R_2}{R-R_1}\right)^{2/3}, \
\left(\frac{R-R_1}{R-R_2}\right)^{1/3}, \ \left(\frac{R-R_1}{R-R_2}\right)^{1/3}\right]
\, .
\ee
Let us now set $C_1 C_2=3$. The components of the metric are determined by computing
$\Sigma= - X^{-1/2} F$, where we need to choose $\sigma=-1$ to get the canonical
expressions for the ASD 2-forms $\Sigma^i$. We then get the frame fields
\ber
&&\theta^t = \frac{C_1}{\sqrt{C_2}} \sqrt{1+C_2(R-R_2)^{2/3}} dt, \quad \theta^R=
\frac{\sqrt{C_2}dR}{3\sqrt{1+C_2(R-R_2)^{2/3}} (R-R_1)^{1/3}(R-R_2)^{1/3}}, \nonumber \\
\\  &&\theta^\theta = \sqrt{C_2} (R-R_1)^{1/6}(R-R_2)^{1/6} d\theta, \quad \theta^\phi =
\sqrt{C_2} (R-R_1)^{1/6}(R-R_2)^{1/6} \sin \theta\, d\phi \, ,
\eer
and so the metric reads
\ber\label{ss-metric}
ds^2_4 &=& \frac{C_1^2}{C_2}(1+C_2(R-R_2)^{2/3}) dt^2 + \frac{C_2
dR^2}{9(1+C_2(R-R_2)^{2/3})(R-R_1)^{2/3}(R-R_2)^{2/3}} \nonumber \\
&&{}+ C_2 (R-R_1)^{1/3}(R-R_2)^{1/3} d\Omega^2\, ,
\eer
where, as usual, $d\Omega^2$ is the unit sphere metric. Asymptotically for large $R$,
introducing $\sqrt{C_2} R^{1/3}=r$ we get the following metric: $ds^2_4 = r^2
\left[(C_1^2/C_2) dt^2 + d\Omega^2 \right] + dr^2/r^2$. This is the hyperbolic space
metric with the conformal structure of the boundary being that of $S^1\times S^2$,
provided we identify the `time' coordinate $t$ periodically. It is now easy to write the
lift (\ref{metric-om}) of the metric (\ref{ss-metric}). We get
\ber
ds^2_7 &=& \left( 1 - y^2 \right)^{-1/2} \left[ \left(\frac{R-R_2}{R-R_1}\right)^{2/3}
\left[ dy^1 - b (R) \sin \theta\, d\phi \, y^3 - b(R) d\theta \, y^2 \right]^2 \right. \nonumber \\
&&{}+ \left(\frac{R-R_1}{R-R_2}\right)^{1/3} \left[ dy^2 +  b(R) d\theta \, y^1 - (a(R)dt
+\cos \theta\,  d\phi) y^3 \right]^2 \nonumber \\
&&{}+ \left. \left(\frac{R-R_1}{R-R_2}\right)^{1/3} \left[ dy^3 +( a(R)dt +\cos \theta\,
d\phi) y^2 + b(R) \sin \theta\, d\phi \, y^1 \right]^2 \right] \nonumber \\
&&{}+ 2 \left( 1 - y^2 \right)^{1/2} ds_4^2\, ,
\eer
with $a(R)$ and $b(R)$ given by (\ref{ab}).

\subsection{Bianchi IX}

As we already mentioned, our treatment of this case is incomplete, because we are unable to solve the arising ODE's. The main result of this subsection is the ODE (\ref{bi-axial-eq}) to which the problem reduces in the bi-axial case. If one can solve this ODE (e.g. numerically) one would obtain cohomogeneity one 4D metrics that asymptote to Bianchi IX bi-axial instantons -- the Taub-NUT metrics. 

\subsubsection{Ansatz}

Let $e^1$, $e^2$, $e^3$ be the standard basis of $\su (2)$ left-invariant one-forms on
$S^3$ with structure equations $d e^1 = e^2 \wedge e^3$ etc.  Consider the ansatz
\beq
A^1 = h_1 e^1 \, , \qquad A^2 = h_2 e^2 \, , \qquad A^3 = h_3 e^3 \, , \qquad
\eeq
where $h_i$ are functions of the `radial' coordinate $r$.  The curvature components are
\beq
F^1 = h_1' d r \wedge e^1 + \left( h_1 + h_2 h_3 \right) e^2 \wedge e^3 \, , \quad
\mbox{etc} \, .
\eeq

\subsubsection{The metric}

Our first aim is to calculate the appropriately normalized (Euclidean) metric in which
the curvature components are anti-self-dual.  As before, we do this computation by
computing the matrix of curvature wedge products $F^i\wedge F^j = 2X^{ij} dr\wedge
e^1\wedge e^2\wedge e^3$:
\be
X = {\rm diag} \left[ h_1'(h_1+h_2 h_3), h_2'(h_2+h_1h_3), h_3'(h_3+h_1h_2)\right] \, .
\ee
We assume all $h_i$ and their derivatives $h_i'$ to be positive, so that the connection
is definite.  We then get $\Sigma = - X^{-1/2} F$, and read off the basis of frame
1-forms by comparing the resulting $\Sigma$'s with (\ref{sigmas}). Note that the sign of
the connection must be chosen to be $\sigma=-1$ to identify the curvature 2-forms with
ASD forms. We get the metric of the form
\beq
d s^2 = N^2 \left( d r \right)^2 + \sum_{i=1}^3 a_i^2 \left( e^i \right)^2\, ,
\eeq
with
\beq \label{lapse}
N^2 = \frac{\left( h_1' h_2' h_3' \right)^{2/3}}{\left[ \left( h_1 + h_2 h_3 \right)
\left( h_2 + h_3 h_1 \right) \left( h_3 + h_1 h_2 \right) \right]^{1/3}} \, ,
\eeq
\beq \label{sfac}
a_1^2 = \frac{\left( h_1' \right)^{2/3}}{\left( h_2' h_3' \right)^{1/3}} \frac{\left[
\left( h_2 + h_3 h_1 \right) \left( h_3 + h_1 h_2 \right) \right]^{2/3}}{\left( h_1 + h_2
h_3 \right)^{1/3}} \, , \quad \mbox{etc} \, .
\eeq

\subsubsection{The action}

The action of the theory is
\beq \label{action1}
S = \int f \left( F \wedge F \right) = V_{S^3} \int \Bigl[ h_1' h_2' h_3' \left( h_1 +
h_2 h_3 \right) \left( h_2 + h_3 h_1 \right) \left( h_3 + h_1 h_2 \right) \Bigr]^{1/3} d
r \, ,
\eeq
where $V_{S^3} = \int_{S^3} e^1 \wedge e^2 \wedge e^3$.  Variation of this action with
respect to $h_i$, $i = 1, 2, 3$, gives equations of motion. Choosing the `radial'
coordinate $r$ so that the Lagrangian in \eqref{action1} becomes constant (equal to
unity), we obtain the following system of equations:
\beq \label{eqs1}
\left( \frac{1}{h_1'} \right)' = \frac{1}{h_1 + h_2 h_3} + \frac{h_3}{h_2 + h_3 h_1} +
\frac{h_2}{h_3 + h_1 h_2} \, ,
\eeq
and equations obtained by cyclic transmutation of indices.  A particular solution of
these equations for all positive and increasing $h_i$ asymptotically tends to the
shape-preserving expansion:
\beq \label{asym}
h_1 \propto h_2 \propto h_3 \propto r^{1/3} \quad \mbox{as $r \to \infty$} \, .
\eeq

\subsubsection{Self-dual case}

In the particular self-dual case, characterised by the condition $F^i \wedge F^j \propto
\delta^{ij}$, we have
\beq \label{SD-eq}
h_1' \left( h_1 + h_2 h_3 \right) = h_2' \left( h_2 + h_3 h_1 \right) = h_3' \left( h_3 +
h_1 h_2 \right) = 1 \ \, ,
\eeq
and equations (\ref{eqs1}) are satisfied identically.  As for the metric components
\eqref{lapse} and \eqref{sfac}, we have
\beq
N^2 = h_1' h_2' h_3' \, , \qquad a_1^2 = \frac{h_1'{}^2}{N^2} \, , \quad \mbox{etc} \, .
\eeq

Equations \eqref{SD-eq} imply the relation
\beq \label{relation1}
h_1^2 + h_2^2 + h_3^2 + 2 h_1 h_2 h_3 = 6 r \, ,
\eeq
where a particular shift of the variable $r$ was made to absorb the integration constant.

\subsubsection{Self-dual bi-axial case}

One can find a family of exact solutions with the biaxial ansatz
\beq \label{biaxial}
h_1 = a \, , \qquad h_2 = h_3 = b \, .
\eeq
In this case, equations \eqref{SD-eq} read
\beq
a' (a + b^2) = b b' (1 + a) \, ,
\eeq
and have an integral
\beq \label{b2}
b^2 = k (1 + a)^2 - 2 a - 1 \, ,
\eeq
where $k$ is the integration constant.  Relation \eqref{relation1} then determines the
radial coordinate as a function of $a$\,:
\beq
r = \frac{k (1 + a)^3}{3} - \frac{(1 + a)^2}{2} + \frac{1}{6} \, .
\eeq

The metric components of this solution read
\ber
N^2 d r^2 &=& \frac{k x - 1}{x \left[ 1 + x \left( k x - 2 \right) \right]} d x ^2 \, , \\
\nonumber \\
a_1^2 &=& \frac{x \left[ 1 + x \left( k x - 2 \right) \right]}{k x - 1} \, , \\
a_2^2 = a_3^2 &=& x \left( k x - 1 \right) \, ,
\eer
where we have introduced a new radial variable $x = 1 + a$.

By changing the radial variable from $x$ to $\rho$ as
\beq
x = 2 n^2 \left( 1 \mp \frac{\rho}{n} \right)\, , \qquad n = \frac{1}{2 \sqrt{k}} \, ,
\eeq
we bring the metric into the form presented in \cite{Cvetic:2002kj}:
\beq
d s^2 = \frac{\rho^2 - n^2}{\Delta} d \rho^2 + \frac{4 n^2 \Delta}{\rho^2 - n^2} \left(
e^1 \right)^2 + \left( \rho^2 - n^2 \right) \left[ \left( e^2 \right)^2 + \left( e^3
\right)^2 \right] \, ,
\eeq
where
\beq
\Delta = (\rho \mp n)^2 \left[ 1 + (\rho \pm 3 n) (\rho \mp n) \right] \, .
\eeq
It describes the self-dual (upper sign) or anti-self-dual (lower sign)
Taub--NUT--anti-de~Sitter metric.

\subsubsection{General bi-axial case}

In the general case with biaxial ansatz \eqref{biaxial}, equations of motion \eqref{eqs1}
read
\beq
\left( \frac{1}{a'} \right)' = \frac{1}{a + b^2} + \frac{2}{1 + a} \, , \qquad \left(
\frac{1}{b'} \right)' = \frac{1}{b} + \frac{b}{a + b^2}\, .
\eeq
It is convenient to change the variables to
\beq
x = 1 + a \, , \qquad y = b^2 - 1 \, ,
\eeq
so that these equations become
\beq \label{xy}
\left( \frac{1}{x'} \right)' = \frac{1}{y + x} + \frac{2}{x} \, , \qquad \left(
\frac{1}{y'} \right)' = \frac{1}{2 (y + x)} \, .
\eeq

We can obtain a closed differential equation for the trajectory $y (x)$.  We have
\beq
\frac{d^2 y}{d x^2} = \frac{1}{x'} \left( \frac{y'}{x'} \right)' = \frac{y''}{x'{}^2} +
\left( \frac{1}{x'} \right)' \frac{y'}{x'} = - \left( \frac{1}{y'} \right)' \left(
\frac{y'}{x'} \right)^2 +  \left( \frac{1}{x'} \right)' \frac{y'}{x'} \, .
\eeq
Now, using equations \eqref{xy}, we get
\beq\label{bi-axial-eq}
\frac{d^2 y}{d x^2} + \frac{1}{2 (y + x)} \left( \frac{d y}{d x} \right)^2 - \left(
\frac{1}{y + x} + \frac{2}{x} \right) \frac{d y}{d x} = 0 \, .
\eeq
This equation looks very difficult to solve.  Note that its partial solution $y (x) = k
x^2 - 2 x$ is precisely the self-dual solution \eqref{b2}.

We also have the constraint reflecting our choice of radial variable:
\beq
x' y'{}^2 (y + x) x^2 = 4 \, .
\eeq
It is only needed to determine the original radial variable $r$ along the trajectory:
\beq
d r = \left[ \frac14 \left(\frac{d y}{d x} \right)^2 (y + x) x^2 \right]^{1/3} d x \,.
\eeq
While we are unable to solve the general bi-axial case analytically (apart from the already known self-dual case), the ODE (\ref{bi-axial-eq}) can be used for a numerical solution, which can then be lifted to 7D.

\section{Discussion}

In this paper we generalised the construction of \cite{BS} by parametrising the 3-form in 7 dimensions with an ${\rm SO}(3)$ connection on the 4-dimensional base instead of a self-dual Einstein metric. We get a $G_2$ holonomy metric in 7 dimensions provided the connection satisfies the Euler-Lagrange equations of the theory (\ref{f*}). Our construction then intersects with that of Bryant-Salamon precisely for self-dual Einstein metrics -- instantons. These are also the solutions that are shared by the theory (\ref{f*}) and GR. The solutions of (\ref{f*}) that are not instantons give metrics on the base that are not Einstein. Our work interprets these non-Einstein metrics as restrictions of 7 dimensional Ricci flat metrics to the 4-dimensional base. 

Our construction shows that a certain theory of gravity in 4D can be understood as
arising via a dimensional reduction of a theory of differential forms in 7 dimensions.
This realisation of a 4D gravity theory as coming from a theory of a very different
nature appears to be the most interesting aspect of our work. While the proper interpretation of our 
dimensional reduction is still to be developed, we give some comments in this direction, leaving the final word to
future studies.

If we are to interpret a 4D gravity as arising by some sort of Kaluza--Klein reduction
from a higher-dimensional theory of differential forms, the must first answer the question 
what this theory of forms is. As we have already described in the main text, a
(stable) 3-form in 7 dimensions naturally defines a metric, and one can compute the
volume of the manifold with respect to this metric. This leads to functional
(\ref{review-func}) whose critical points are 3-forms that are co-closed $d{}^*\Omega=0$,
provided one varies the 3-form within a fixed cohomology class $\Omega = \Omega_0 + dB$,
$B\in \Lambda^2 E$. Thus, one possible interpretation of theory (\ref{review-func}) is as
a theory of 2-forms $B$ in 7 dimensions, with the action constructed from their field
strength $\Omega = dB$. Viewed in this light, it becomes an analog of Maxwell's theory,
where the basic field is a 1-form $A$, and the action is constructed from the field
strength $F=dA$. The principal difference between these two cases is that Maxwell's
theory requires a metric for its formulation, while the theory of 2-forms in 7 dimensions
exists on an arbitrary differentiable manifold without any extra structure. Moreover, the
field strength $\Omega=dB$ itself {\it defines\/} a metric on $E$, and this is why such a
theory can ultimately be given, as in our construction, some gravitational
interpretation. Considering this theory on classes of forms of type $\Omega = \Omega_0 +
dB$ with fixed non-trivial $\Omega_0$, one can interpret this as a theory around
different vacua, with different cohomology classes to be summed over in the path
integral.

Having defined the 7-dimensional theory as a diffeomorphism-invariant analog of
electromagnetism in 7 dimensions (but now for 2-forms rather than one-forms), we can
discuss the meaning of our ansatz (\ref{our-3}) in which we parametrised $\Omega$ by a
connection 1-form on the 4-dimensional base. Our first remark is that, because our ansatz
(\ref{our-3}) is closed, it can locally be written as $\Omega = dB$, with some $B$
parametrised by the connection field. This way of writing $\Omega$ is of course not
unique, because $B$ is only defined modulo $B\to B+d\theta$, $\theta\in \Lambda^1 E$. We
can now get some insight into the meaning of our ansatz (\ref{our-3}) if we consider the
manifold $E$ to have the structure of a product manifold with 3-dimensional fibers and a
4-dimensional base. Then the $21$ components of $B$ decompose as follows: we get 3
components of $B$ that is a 2-form in the fiber, these should be interpreted as scalars
from the point of view of the base; we get the components of $B$ that are basic 1-forms,
as well as 1-forms in the fiber direction, these are interpreted as three 1-forms on the
base; finally, there is the component that is a 2-form on the base. Counting the numbers
we get $3+ 3\times 4 + 6=21$ components as required.

It is then clear that if we make an assumption that $B$ is invariant under an action of
some group in the fiber directions, we get a structure of a fiber bundle with the basic
1-form components of $B$ receiving the interpretation of a connection in this bundle. In
our ansatz (\ref{our-3}), we have parametrised $\Omega$ and thus $B$ just by these
connection components, thus setting to zero all other possible fields that could have
been present. This led us to a diffeomorphism-invariant ${\rm SO}(3)$ gauge theory with
the Lagrangian $( \det  X )^{1/3}$ on the base. It would be very interesting to keep all
the components of the 2-form $B$, and find the resulting theory. This is to be described
elsewhere.

It thus appears that the right interpretation of our construction is that we have made a
particular Kaluza--Klein ansatz (\ref{our-3}) for some 7-dimensional theory, and saw how
the 7-dimensional field equations impose some 4D field equations on our ansatz. This is
not yet Kaluza--Klein reduction, in which one would instead make an assumption that the
7-dimensional field is invariant with respect to some group action, and determine all the
fields and their dynamics that arise in lower dimensions. To put it differently, what we
have obtained is analogous to the Kaluza--Klein ansatz that obtains gravity plus Maxwell
in 4 dimensions from gravity in 5 dimensions, while setting the other field necessarily
present in this dimensional reduction --- the scalar field --- to zero by hand. It is
very interesting to determine the full content and dynamics of the 7D theory
(\ref{review-func}) dimensionally reduced to 4D.

A potentially more difficult question is to study the dimensional reduction without any
symmetry assumptions, and thus take into account the full infinite set of modes that
arise by decomposing the field dependence on the `internal' coordinates into appropriate
spherical harmonics. In the usual Kaluza--Klein story, the fields with non-trivial
dependence on the internal coordinates get interpreted as massive modes. It would be very
interesting to see if the same interpretation persists for the 7-dimensional theory of
differential forms.

Finally, the most interesting physics question that arises in this context is whether
General Relativity (\ref{f-GR}) can arise by a similar dimensional reduction from a
higher-dimensional theory of differential forms. This could be precisely GR in its form (\ref{f-GR}), or, perhaps,
a theory that only resembles GR in some appropriate range of energies,
but differs from it in general. At the moment of writing this, it appears to us that
this last possibility is the most likely one. 

Whatever the final word on this story will be, it appears clear to us that there is a
very interesting class of dynamically non-trivial theories of differential forms in
higher dimensions, e.g., 3-forms in 7 and 8 dimensions, see \cite{Hitchin} for details of
how the action functional is defined in 8D. The construction presented in this paper
makes it clear that these theories, when dimensionally reduced, are related to
dynamically non-trivial gravity theories in 4 dimensions. The main question now is what
gravity theories can arise in this way, and what kind of other fields that accompany
gravity arise in such a dimensional reduction. Our paper can be viewed as a first step in
answering these questions.

\section*{Acknowledgments}

KK and CS were supported by ERC Starting Grant 277570-DIGT\@. YS acknowledges support
from the same grant and from the State Fund for Fundamental Research of Ukraine. 
YH was supported by a grant from ENS Lyon. KK is grateful to Joel Fine for
important discussions. The idea that later developed into this paper was born in one of
them.

\end{document}